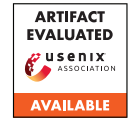 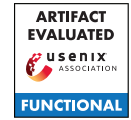 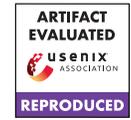

# IvySyn: Automated Vulnerability Discovery in Deep Learning Frameworks

Neophytos Christou
*Brown University*

Di Jin
*Brown University*

Vaggelis Atlidakis
*Brown University*

Baishakhi Ray
*Columbia University*

Vasileios P. Kemerlis
*Brown University*

## Abstract

We present `IvySyn`, the first fully-automated framework for discovering memory error vulnerabilities in Deep Learning (DL) frameworks. `IvySyn` leverages the statically-typed nature of native APIs in order to automatically perform type-aware mutation-based fuzzing on low-level kernel code. Given a set of offending inputs that trigger memory safety (and runtime) errors in low-level, native DL (C/C++) code, `IvySyn` automatically synthesizes code snippets in high-level languages (e.g., in Python), which propagate error-triggering input via high(er)-level APIs. Such code snippets essentially act as *"Proof of Vulnerability"*, as they demonstrate the existence of bugs in native code that an attacker can target through various high-level APIs. Our evaluation shows that `IvySyn` significantly outperforms past approaches, both in terms of efficiency and effectiveness, in finding vulnerabilities in popular DL frameworks. Specifically, we used `IvySyn` to test TensorFlow and PyTorch. Although still an early prototype, `IvySyn` has already helped the TensorFlow and PyTorch framework developers to identify and fix 61 previously-unknown security vulnerabilities, and assign 39 unique CVEs.

## 1 Introduction

Artificial Intelligence (AI) is increasingly employed by software systems that impact virtually every aspect of modern society, ranging from economy and health to science and education. In fact, the US Department of Energy predicts that during the next five years, AI will be part of *mission-critical* systems that affect the health, safety, and welfare of the public, such as systems related to: our telco. infrastructure; water supply and electrical power; roadway, railway, and air transportation; banking and financial services; public safety; healthcare; as well as public services and administration [38]. Deep Learning (DL) has been at the forefront of such efforts. For instance, mobility-as-a-service providers, like Uber, rely on DL to match riders to drivers, suggest optimal routes, find pool combinations, and create next-generation intelligent vehicles, in an attempt to provide reliable transportation [66].

PyTorch [46] and TensorFlow [1] are perhaps the most widely-adopted DL frameworks that are used in various *safety-critical* settings, like autonomous driving: e.g., Lyft and Tesla rely on PyTorch to solve problems related to the self-driving domain, such as mapping, perception, prediction, and planning [50, 76]. In addition, TensorFlow is part of the operational infrastructure of key stakeholders in technology, including Google, Twitter, IBM, Intel, Arm, and Qualcomm [59].

Such frameworks typically provide a set of rich high-level APIs to model developers, to allow them integrate DL functionality into end-to-end, real-world systems. These developer-accessible APIs perform DL-specific *operations*, and are usually implemented in managed languages, like Python. Yet, the essential parts of such operations—which are called *kernels*—are implemented in memory-unsafe languages, like C/C++, to boost performance and to allow for execution in different platforms (e.g., CPUs, GPUs, TPUs). DL frameworks have bugs in their underlying implementations, and it has been empirically shown that such bugs may cause incorrect functionality, numerical errors, performance degradation, memory exhaustion and CPU/GPU/TPU lockups, and fatal runtime and memory safety errors [28, 31, 44, 77].

Bugs in *kernel code* are of special interest, however, as they may result in memory safety errors [67] that can, in turn, be abused by attackers for corrupting or leaking memory contents [58], or even causing the respective runtime environment to crash, effectively leading to a DoS. The latter can be extra problematic in cloud settings (e.g., in AWS DL Containers [2] or IBM Watson Discovery [30]) because an attacker with access to publicly-available, high-level APIs can send requests with specially-crafted inputs that exploit memory-safety errors in low-level kernel code [29].

In 2021–2022 alone, TensorFlow had more than 280 CVE numbers assigned for (potentially-exploitable) vulnerabilities related to memory safety issues [14]. (Similar vulnerabilities also exist in the PyTorch codebase.) Given that, we ask the following questions: Can we detect such vulnerabilities—affecting millions, if not billions, of users—in an *automated* and actionable manner?

Even across complex frameworks with multiple layers of abstractions, implemented in codebases consisting of different low- and high-level languages, with potentially mismatched security assumptions?

Although past approaches have attempted to answer similar questions by employing fuzzing directly on high-level APIs, they are either: semi-automated, and require *domain-expert annotations* for specifying valid argument-value combinations [74]; or not-automated at all, and require developers to manually write *helper code* [23]. We take the converse path and introduce a two-fold bottom-up approach, implemented in our `IvySyn` framework. `IvySyn` leverages the statically-typed nature of native APIs in order to automatically perform type-aware mutation-based fuzzing on low-level kernels. Next, given a set of offending inputs that trigger memory safety (and fatal runtime) errors in low-level, native DL (C/C++) code, it automatically synthesizes code snippets in managed languages (e.g., Python), which propagate offending input through high(er)-level APIs. Such code snippets essentially act as *"Proof of Vulnerability"* (PoV), as they demonstrate the existence of bugs in native, C/C++ code that attackers can target (and potentially abuse) via high-level APIs.

We implemented `IvySyn` in C/C++ and Python (≈3 KLOC), and our experimental evaluation shows that `IvySyn` significantly outperforms past approaches, both in terms of efficiency and effectiveness, in finding real vulnerabilities in popular DL frameworks. Specifically, we used `IvySyn` to test TensorFlow and PyTorch: although still an early research prototype, `IvySyn` has already helped the TensorFlow and PyTorch framework developers to identify and patch 61 previously-unknown security vulnerabilities, in the latest production versions, and assign 39 unique CVEs.

## 2 Background and Motivation

**Typical Architecture of Deep Learning Frameworks.** DL frameworks, such as TensorFlow [1] and PyTorch [46], typically consist of multiple layers of abstractions. At the lowest layer, native *low-level APIs*, or *kernels*, implement DL-specific operations for various devices, such TPUs, GPUs, and CPUs. Depending on the supported devices, each *operation* may have one, or more, equivalent kernels available for it. Operations that are intended to be accessible by model developers have *bindings* mapping them to high-level languages (e.g., Python), while, ultimately, these bindings are wrapped around *high-level APIs* that expose them to framework users.

• *Kernels*: The core functionality of DL frameworks is provided by native C/C++ code, implementing standard operations, such as tensor manipulation, mathematical, convolutional, and gradient computations, pooling, and other DL-specific operations. The actual, native implementations of such operations are called *kernels*. Implementing core operations in a language like C/C++ not only provides better performance, but also allows framework developers to implement multiple, optimized versions for different hardware devices (e.g., CPUs, GPUs, and TPUs). The kernels are recommended to avoid using (or depending on) *shared state* [61], with each invocation being self-contained, in order to be easily parallelizable. This is an important property that `IvySyn` builds upon to seamlessly add its kernel fuzzing hooks (§5).

• *Bindings*: Even though DL frameworks provide low-level, kernel APIs, model developers typically perform DL-specific operations through high-level languages, such as Python. Thus, in order to expose a subset of these operations through developer-accessible APIs, DL frameworks generate high-level *bindings* [52, 61]. These bindings translate input arguments fed to, say, Python APIs, into the corresponding C/C++ arguments, and transparently invoke the appropriate kernel, abstracting away implementation details regarding the underlying foreign-function interface. `IvySyn` builds upon DL framework mappings (bindings ⇄ kernels) and low-level crashing inputs to seamlessly synthesize code snippets involving the respective high-level APIs (§5.4). Even though these bindings are not necessarily intended to be used by developers, they are still publicly-exposed and callable from, say, Python, giving an attacker direct access to exploitable vulnerabilities in the native kernel implementations.

• *High-level APIs and DL Models*: High-level language bindings (such as the ones described above) are further wrapped by other, high(er)-level APIs that are language-specific (e.g., implemented in Python). The latter may add some abstraction over the raw binding, like pre-processing or extra sanity checks for the arguments, before invoking the respective operations; or may be mere wrappers, adding documentation and uniformly-exporting operations to framework modules. Eventually, the high-level APIs that wrap around bindings of kernel operations are the ones intended to be used by model developers. We call these developer-accessible and documented APIs, *high-level APIs*. Typically, multiple such high-level APIs are utilized when building *DL models* for various tasks, such as image classification for self-driving cars [36], malware detection [54], and face recognition [17].

**Testing Deep Learning Frameworks.** DL frameworks have bugs in their underlying implementations, and it has been empirically shown that such bugs may cause memory safety errors, fatal runtime errors, incorrect or inconsistent functionality, numerical errors, performance degradation, memory exhaustion and CPU/GPU/TPU lockups [28, 31, 44, 77]. Bugs in the native, C/C++, parts of DL frameworks (e.g., kernels) are of special interest, since they may result in fatal runtime errors and memory errors that can be abused by attackers.

One path towards finding such bugs is to make use of pre-trained DL models and continuously: (1) perform alternations on them [26, 44, 71, 72, 75]; and (2) feed them back to the target DL frameworks hoping to trigger erroneous behaviors. However, since the unit of mutations is DL models, these works mostly uncover errors related to incorrect/inconsistent functionality, numerical errors, and degradation in accuracy.

On the other hand, mutation-based approaches that perform alternations on the input arguments of high-level APIs [23,74], indeed generate erroneous inputs that transfer through the stack of abstractions to native kernel implementations, and trigger memory safety and fatal runtime errors.

However, because high-level APIs are not statically-typed, past approaches are semi-automated or completely manual: a domain-expert has to *manually* add annotations (with precise information about valid argument-value combinations) [74] or implement helper code (i.e., drivers) and add type-awareness [23] for high-level APIs.

**Automated Vulnerability Discovery.** Our goal is to *automatically* uncover offending inputs that are passed to high-level DL framework APIs, transfer through the various layers of abstractions, and trigger fatal runtime and memory errors in low-level, native code. Our key insight towards full automation, is to take a two-fold, bottom-up approach: first, we leverage the statically-typed nature of native APIs, which allows us to perform type-aware mutation-based fuzzing on low-level kernel code; next, given a set of offending inputs for low-level APIs, we leverage the inherent DL framework mappings from low- to high-level APIs and synthesize code snippets, which trigger memory errors on low-level, native code with inputs passed to high-level APIs. We realize our bottom-up approach in a framework called IvySyn, which needs no domain-expert annotations nor manually-written driver code.

## 3 Threat Model and Memory Errors

**Adversarial Capabilities.** We assume an attacker who is aiming at *exploiting* DL frameworks (e.g., TensorFlow [1], PyTorch [46]) by triggering *memory errors* [67] in their respective kernel codebases. Specifically, we focus on attackers who (ab)use *spatial* [42] or *temporal* [43] memory safety violations, in DL framework components written in type- and/or memory-unsafe languages (e.g., C, C++), with the end goal of tampering-with critical *code/data pointers* [48].

Platforms like IBM's Watson Discovery [30] are assumed to be the canonical example of a victim, "wrapper" application that internally leverages a DL framework (i.e., TensorFlow), which contains memory error-based vulnerabilities that can be further triggered via bogus input to the wrapper part. In the case of Watson Discovery, any TensorFlow memory error (re)classified by IBM as a vulnerability with the following description *"by sending a specially-crafted request, an attacker could exploit this vulnerability to ..."* [29] is a prime example of our focus. By identifying vulnerabilities in memory-unsafe framework code, IvySyn aids framework developers in fixing the respective bugs (§7.3), thereby implicitly protecting any application that uses the corresponding DL framework as well, via means of removing vulnerabilities from a perceived exploitation path from the wrapper application (e.g., Watson Discovery) to the low-level, native framework code.

**Memory Errors.** IvySyn aims at uncovering crashing inputs that trigger: (a) *memory safety errors*, like (arbitrary) memory corruption and memory disclosure vulnerabilities [58]; and (b) *fatal runtime errors*—both (a) and (b) manifest as crashes (i.e., abnormal process termination with SIGABRT, SIGFPE, and SIGSEGV exceptions) during fuzzing/runtime.

Note that IvySyn is a software testing framework, and similarly to the fact that tools that discover memory errors in "library" code [35, 55] do not necessarily study how the uncovered bugs can be exploited within the context of the applications that use them, it does not provide automation regarding the end-to-end exploitation of the uncovered DL-framework vulnerabilities. However, IvySyn's PoVs (§5.4) provide useful information (offending arguments, vulnerable APIs, *etc.*) to form an attack path (see Appendix B.)

## 4 Approach Overview

We introduce a *two-fold*, *bottom-up* approach for testing DL framework implementations with the end goal of uncovering security vulnerabilities. We realize our bottom-up approach in a framework called IvySyn, which, given a set of native DL kernels, aims at answering two questions *automatically*:

1. Are there any offending inputs to low-level APIs that can trigger memory errors in native DL kernels?

2. Are there any high-level APIs that can propagate the above offending inputs to low-level kernel code?

To uncover offending inputs that can trigger memory errors in native DL kernel implementations, IvySyn leverages mutation-based fuzzing [21], augmented with *type-aware* mutations (§5.3). Then, armed with a set crashing inputs, IvySyn automatically *synthesizes* appropriate code snippets that trigger the respective errors from high-level APIs.

Every such code snippet acts as a *"Proof of Vulnerability"* (PoV), as it demonstrates: (1) *the existence of a bug in kernel* code that can result in memory corruption or disclosure (integrity/confidentiality violation), or even halt the whole DL framework runtime (availability violation); and (2) *that an attacker can trigger this bug via a high-level API*, which is available to model developers or even end users (depending on how the DL model "consumes" untrusted user inputs when integrated into real-world systems) [29].

**Crash Reports.** Listing 1 illustrates a crash report that was generated by IvySyn while fuzzing the kernel Edit-DistanceOp of the TensorFlow API tf.raw_ops.Edit-Distance—which is used for computing the Levenshtein edit distance. The report captures the offending input(s) of a *real*, previously-unknown, vulnerability that was uncovered by IvySyn, and corresponds to an *invalid memory write* via a pointer (i.e., memory corruption). IvySyn crash reports contain information regarding the offending input, including the concrete values of the tensor(s) involved (ln. 2–9).

```
1  # EditDistanceOp::Compute()
2  Tensor<type: int64 shape: [3,3]
3          values: [-1250999896764 ...]...>
4  Tensor<type: int64 shape: [3] values: 0 0 0>
5  Tensor<type: int64 shape: [3] values: 0 0 0>
6  Tensor<type: int64 shape: [2,3] values: [0 0 0]...>
7  Tensor<type: int64 shape: [2]
8          values: -1879048192 -1879048192>
9  Tensor<type: int64 shape: [3] values: 2 2 2>
```

Listing 1: Crash report produced by IvySyn while fuzzing the kernel EditDistanceOp::Compute of the TensorFlow API tf.raw_ops.EditDistance.

```
1  import tensorflow as tf
2
3  hypothesis_indices = tf.constant(-125099896764,
4                       shape=[3,3], dtype=tf.int64)
5  hypothesis_values  = tf.constant(0,
6                       shape=[3], dtype=tf.int64)
7  hypothesis_shape   = tf.constant(0,
8                       shape=[3], dtype=tf.int64)
9  truth_indices      = tf.constant(0,
10                      shape=[2,3], dtype=tf.int64)
11 truth_values       = tf.constant(-1879048192,
12                      shape=[2], dtype=tf.int64)
13 truth_shape        = tf.constant(2,
14                      shape=[3], dtype=tf.int64)
15
16 tf.raw_ops.EditDistance(
17     hypothesis_indices = hypothesis_indices,
18     hypothesis_values  = hypothesis_values,
19     hypothesis_shape   = hypothesis_shape,
20     truth_indices      = truth_indices,
21     truth_values       = truth_values,
22     truth_shape        = truth_shape)
```

Listing 2: PoV synthesized by IvySyn for triggering the vulnerability of Listing 1 (TensorFlow kernel EditDistanceOp::Compute via binding tf.raw_ops.EditDistance).

Since IvySyn produces crash reports automatically, one can synthesize C/C++ programs that invoke a crashing kernel with the respective offending input. However, a native code snippet would not be sufficient for framework developers to replicate and analyze the corresponding bug: it is unclear whether the bug is actually triggerable by an API available to model developers (or users). When presented with a crashing kernel, framework developers have to *manually*: (a) identify high-level APIs that, when invoked, result into the execution of the offending kernel code; (b) experiment with the arguments of those (high-level) APIs so that the target (low-level, kernel) code ends-up being invoked with bogus input; and (c) put together a Python snippet that demonstrates the issue end-to-end. IvySyn completely automates (a), (b), and (c); given a crash report in kernel code, it synthesizes end-to-end running *programs* (PoVs), which demonstrate the existence of a bug that can by triggered via high-level APIs.

**PoVs.** Listing 2 shows an IvySyn PoV that triggers the previously-identified vulnerability in the C++ *kernel* EditDistanceOp, by using the high-level API tf.raw_ops.EditDistance (ln. 16–22). The resulting PoV is enough to demonstrate the existence of a bug in kernel code that can by triggered via a high-level API: one just needs to merely execute the PoV code. Framework developers can use such PoVs to replicate, analyze and study, and eventually fix the respective bug(s). IvySyn' s PoVs have helped uncover multiple, previously-unknown vulnerabilities in TensorFlow/PyTorch, and have led to multiple bug-fixes and CVEs (§7.3).

## 5 The IvySyn Framework

We designed IvySyn in accordance to three design principles:

1. We follow developer directives about extending DL frameworks in order to (a) identify native APIs and (b) add the appropriate fuzzing hooks to them. (*Insight:* Following such directives allows us to automatically identify native code, while the strongly-typed nature of native APIs provides type-awareness for free.)

2. We hook kernels without shared state and use developer tests to force-execute and fuzz them. (*Insight:* Force-executed kernel fuzzing allows us to invoke native APIs within a proper calling context.)

3. We leverage the inherent DL framework mappings between high- and low-level APIs in order to synthesize PoVs. (*Insight:* Given a set of low-level crashing inputs, working backwards allows to synthesize high-level code snippets almost instantly.)

Figure 1 illustrates the overall architecture of IvySyn, in terms of major components and their interactions. IvySyn's instrumentation extracts native kernel implementations from the respective DL codebase in order to construct and inject fuzzing wrappers (§5.1). IvySyn's Watchdog invokes the entry-points of developer test suites; as soon as the flow of execution reaches a target kernel, IvySyn's instrumentation will bootstrap a force-executed mutation-based fuzzing session (§5.2) using type-aware mutations (§5.3). IvySyn's synthesizer uses low-level crashing inputs uncovered during fuzzing to generate PoVs, which, in turn, trigger low-level memory errors from publicly-available APIs (§5.4).

### 5.1 Instrumentation

To test native code, IvySyn first needs to extract the functions that correspond to DL-framework kernels, and then create and inject appropriate fuzzing wrappers around them.

**Extracting Kernel Implementations.** The process of identifying kernel code is slightly different between PyTorch and TensorFlow. Nevertheless, it is completely *automated* in both frameworks, as `IvySyn` leverages *developer directives* regarding extending framework code. In TensorFlow, each kernel *inherits* from the `OpKernel` interface and implements a `Compute()` method [61]—the body of `Compute()` basically includes the implementation of a particular kernel. `IvySyn` parses the source code of TensorFlow and creates a set of C++ classes that directly inherit from `OpKernel`, or from a class that (directly or indirectly) inherits from `OpKernel`, and overrides `Compute()`. Next, it further prunes that set of classes by removing: (1) kernel implementations that deal with shared state (e.g., include synchronization locks, like `std::tensorflow`, or interact with `ResourceMgr` objects); and (2) other edge cases, such as classes that modify the type signature of `Compute()` (e.g., by altering the arity of the method) or classes that re-define `Compute()` so that it has an empty body.

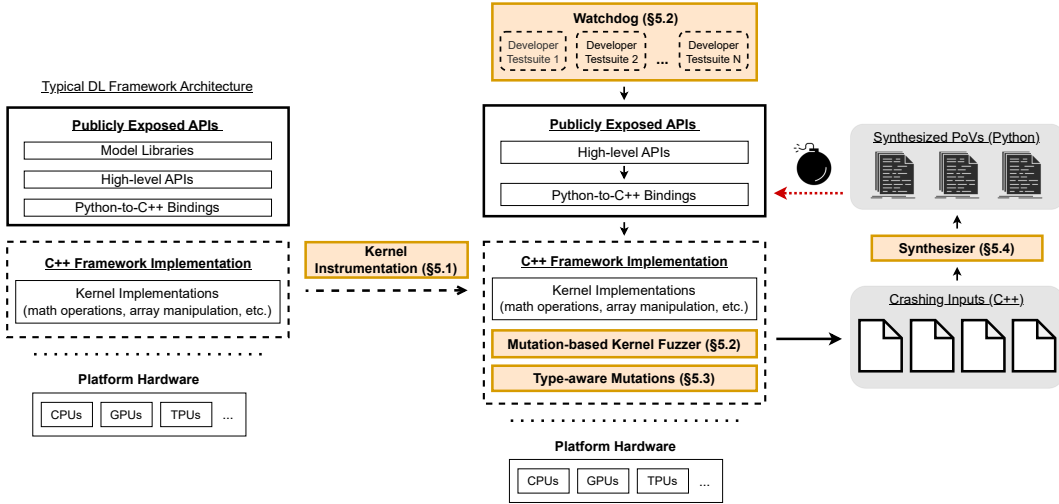

Figure 1: Overview of the `IvySyn` fuzz testing framework.

In PyTorch, each kernel implementation has a corresponding descriptor in the YAML file 'native_functions.yaml' [51]. Each descriptor provides the *symbol name* of the C/C++ function implementation, followed by a list of *arguments* (with their respective symbol names), along with their *types*, and the type of the *return value*. `IvySyn` parses `native_functions.yaml` and creates an initial set of C/C++ symbols that correspond to native code that implements PyTorch kernels. Next, it curates the set by removing symbols in the following categories: (1) `out`-variants of functions—if `f(...)` has a companion function `f.out(..., *rv)`, then we can safely ignore the latter, as it is a version of `f()` where the return value is passed by reference to `rv`; (2) pure wrappers/helpers of functions, merely wrapping-around target kernel(s); and (3) functions with argument types that `IvySyn` does not handle—these correspond to esoteric, non-{tensor, array, scalar} types, like `MemoryFormat` and `Generator`, as well as lists of objects and object indexers. Note that (3) can easily be supported with additional engineering work.

The net effect of all the above is a set of functions that correspond to kernel code for which PyTorch and TensorFlow provide *bindings* to be invoked by high-level APIs and model developers. `IvySyn` creates and injects fuzzing wrappers for these kernels, called *target kernels*.

**Creating and Injecting Wrappers.** Given a set of target kernels, `IvySyn` creates and injects *fuzzing wrappers* as follows: first, it renames the original kernel function, say `func()`, to `do_func()`; then, it replaces the body of the original function with wrapper code, initializing an `IvySyn` fuzzing session.

Listing 3 (in Appendix A) shows a concrete example of `IvySyn`'s instrumentation on the PyTorch kernel `cosine_similarity` that corresponds to the high-level API `torch.nn.functional.cosine_similarity`. Initially, various checks are performed in order to avoid recursive fuzzing or testing kernels that have already been fuzzed (ln. 4–8). In either case, fuzzing is skipped by calling the original kernel implementation. Otherwise, the global variable `already_fuzzing`, indicating that a new fuzzing session has been initiated, is set (ln. 9). Afterwards, the instrumented code passes the original arguments of the target function to the `IvySyn` fuzzer (i.e., the input arguments provided to the original kernel, in the case of PyTorch, or the arguments that are stored in the `context` object, which, in turn, is provided as input to TensorFlow's `Compute`).

The wrapper code also provides an additional vector containing information about the respective argument types (ln. 11–23). This is required because `IvySyn` performs type-aware mutations, and internally the fuzzer uses the original arguments as seeds to create dedicated pools of mutations for each argument type (§5.3). Once the fuzzer has been properly initialized, the main fuzzing loop iterates over the pool of available mutations (ln. 25–37) until all generated mutation combinations are exhausted or a user-defined cap on the maximum number of combinations is reached.

Each combination of input values is used to execute the function under test (ln. 32–33) with the hope of triggering a low-level memory safety or runtime error. Finally, the result of the original invocation is returned (ln. 40–41) so that the invoking code can continue on the normal flow of execution.

### 5.2 Force-executed Kernel Code Fuzzing

DL frameworks typically consist of multiple layers of abstractions, and do not expose low-level APIs (i.e., kernels) for direct usage. Instead, as noted in Section 2, and also illustrated in the left part of Figure 1, kernels are intended to be implicitly used by high-level APIs as follows: high-level APIs ⇝ bindings ⇝ kernels. Therefore, directly fuzzing kernel APIs, without a proper *calling context* (i.e., a specific chain of functions active on the call stack and a set of properly-initialized global variables), will lead to incorrect functionality.

Fortunately, DL frameworks often contain extensive test suites of *developer-provided* unit tests. These are written in high-level languages (e.g., Python) and invoke high-level APIs, which, in turn, call DL framework bindings, and eventually execute the respective kernel code. IvySyn is designed to leverage such developer-provided unit tests and *force-execute* target kernels in order to perform mutation-based fuzzing, with a proper calling context.

To bootstrap a force-executed fuzzing session, IvySyn creates a *Watchdog* process that runs the entry-point of the developer-provided unit tests (e.g., `tf-tests/*.py` in the case of TensorFlow). As explained earlier, during instrumentation, the original kernel function definitions are interposed with IvySyn wrapper code. Therefore, as soon as the execution of a developer test reaches an instrumented kernel, it will kick-start a "force-executed" fuzzing session. This approach works seamlessly because (1) force-executed kernels are invoked with a proper calling context, and (2) kernels usually avoid shared state by design. (Kernels that do interact with shared state are currently ignored by IvySyn; see §5.1.)

**Crash-report Generation.** Each combination of mutated inputs is assigned a *unique index number* (UIN). IvySyn logs the UIN that corresponds to the current mutated input(s) combination into a log file, which is named after the function being fuzzed. Once a crash happens, this log file will contain the UIN of the last attempted input combination, which is the one that caused the memory error. IvySyn leverages this information to re-initialize the fuzzer, get the offending input(s), and stitch together a *crash report* (Listing 1).

**Watchdog.** IvySyn's Watchdog kick-starts a fuzzing session by invoking the entry-points of unit tests, but also monitors all processes performing fuzz testing. Specifically, when a test exits, the Watchdog collects its exit code: a negative code indicates that some native function, which the test invoked, has crashed (abnormal process termination with SIGABRT, SIGFPE, or SIGSEGV). In this case, the Watchdog re-runs the test until the latter exits gracefully—which would mean that the fuzzer has not uncovered any additional crashes for the native functions invoked by the specific test. Furthermore, the Watchdog is responsible for enforcing user-defined timeouts, so that each fuzzing session completes within a desired timeframe, and also prevents potential CPU hogging.

### 5.3 Type-aware Mutations

IvySyn leverages mutation-based fuzzing [21] to stress-test DL kernel code. However, and most importantly, it further augments traditional mutation-based fuzzing with the ability to perform *type-aware* mutations with respect to input types relevant to the target DL framework—such as tensors and arrays. Our intuition is that plain, byte-level mutations (even when guided by code coverage information), will lead to test inputs that hit shallow argument-type checks, without triggering any interesting edge case behaviour. On the contrary, type-aware mutations lead to test inputs that pass argument-type checks, and exercise interesting functionality.

We empirically studied ≈240 CVEs assigned to TensorFlow vulnerabilities, as well as bug reports that involve memory safety issues in PyTorch, and put together a set of mutations for input types relevant to DL operations [34]. Overall, we observed that crashes are commonly caused by inputs that are likely to trigger edge-case behavior, such as tensors with large positive or negative values, empty lists and tensors, and deep tensors with multiple dimensions. Next, we describe an indicative set of mutations currently supported by IvySyn.

- *Tensors and Lists*: For input arguments of type tensor, IvySyn creates a pool of tensor mutations containing tensors with large positive and negative values, tensors with empty shapes, tensors containing random dimensions (including 0-sized dimensions), as well as deep tensors with up to 15 dimensions (configurable limit). Similarly, for list inputs, IvySyn creates a pool of various lists, containing large positive and negative values, as well as empty lists.

- *Concrete Values*: When the input is a single value (e.g., integer, double, boolean, or string), IvySyn creates a pool for the corresponding primitive type, consisting of the zero value, large positive and negative integers and doubles, empty and large strings (e.g., 300-characters long; configurable), *etc*.

- *Original-arguments Permutation*: IvySyn also adds the original, concrete input arguments (passed to the kernel) in the corresponding pool for each type. Hence, when iterating over the mutation pools, some combinations of mutations end-up being permutations of the original arguments.

When initializing the mutation pools for fuzzing a target kernel, IvySyn looks at the types of the input arguments, which are available at the native level, and *constructs a pool of mutation values for each type*, based on the mutation strategies described above. IvySyn then lazily iterates over the pool of all feasible value combinations, until a user-defined timeout is hit or until there are no further mutations to perform.

IvySyn does not use any code coverage information to prioritize some mutations over others—it is completely *black-box* with respect to testing. Yet, as we demonstrate later (§7), IvySyn significantly outperforms state-of-the-art coverage-guided—but type-unaware—approaches, as it finds more crashes, and does so faster. This emphasizes the importance of type-aware mutations and confirms our intuition that byte-level mutations are less successful at uncovering crashes when dealing with complex DL data structures such as tensors.

### 5.4 PoV Synthesis

The ultimate goal of IvySyn is to synthesize PoVs: that is, code snippets in a managed language (e.g., Python) which invoke vulnerable kernel code via a developer-accessible API (e.g., a binding or a high-level API). In order to do so, IvySyn's synthesizer collects crash reports generated during fuzzing and uses them to synthesize appropriate PoVs.

First, given a crash report with information regarding the offending input arguments and the respective crashing kernel implementation (see Listing 1), the synthesizer converts low-level input arguments into the corresponding input arguments for high-level APIs. Specifically, in PyTorch, C++ scalar types, such as `int` and `long`, are converted into a Python `int`; C++ `double` and `float` are converted into Python `float`, *etc.*; and for tensor arguments, the synthesizer constructs the tensor using the information about the tensor's shape, type, contents, and other parameters that were logged into the crash report. In TensorFlow, all arguments are represented as tensors, at the kernel level, and are logged as such. Therefore, the synthesizer, in turn, constructs all arguments as tensors with the proper tensor type, shape and content.

Next, given concrete input arguments for a high-level API, the synthesizer needs to actually identify which high-level API to invoke with them. PyTorch inherently constructs pairs of low-level-function ⇝ binding mapping for each kernel that is exposed to Python. TensorFlow also inherently constructs similar pairs of low-level-function ⇝ binding mappings during framework initialization. IvySyn intercepts those mappings and records them into a file, which is used by the synthesizer to determine the proper binding that corresponds to a low-level crashing kernel. Once the appropriate binding has been identified, the synthesizer creates a call to it and saves the generated PoV (see Listing 2). Lastly, IvySyn iterates through each synthesized PoV, executes it, and categorizes it based on the type of the signal raised after each crash (i.e., `SIGSEGV`, `SIGFPE`, or `SIGABRT`).

Our synthesis approach builds upon the inherent DL framework mappings from bindings to kernels (and vice versa) and is, therefore, straightforward and fully-automated. Furthermore, it is considerably faster than prior work on code synthesis [3, 7]; given low-level crashes, IvySyn takes only seconds to synthesize tens of PoVs (§7).

### 5.5 Design Rationale

We designed IvySyn pursuant to the idea of testing DL frameworks, and selected TensorFlow and PyTorch as our prime targets. We also tried to follow a modular design, without depending on assumptions that fundamentally limit our scope.

More specifically, IvySyn assumes that its target (strongly-typed, low-level) APIs are declared in a well-defined manner (e.g., listed in a YAML file or inherit-from/implement a C++ interface) to instrument them with appropriate fuzzing hooks. In addition, IvySyn assumes the existence of developer-provided unit tests to kick-start kernel fuzzing—usually there is a plethora of unit tests in production-grade software. Since IvySyn is a black-box tool, it requires a set of mutation strategies and input values. To this end, we empirically studied a set of TensorFlow CVEs and PyTorch bug reports, and compiled a preliminary set of mutation strategies. We acknowledge that our approach may be biased towards uncovering bugs particularly in TensorFlow and PyTorch, which were our prime targets. Yet, it is possible to substitute (or extend IvySyn with different) mutation strategies and input values, if deemed appropriate. Finally, IvySyn's synthesis assumes the existence of mappings between high- and low-level APIs. Such mappings are present in any codebase that involves interfacing (safe) managed code with (memory-unsafe) native code.

The core idea behind IvySyn, is its two-fold, bottom-up approach. Extending IvySyn to other DL frameworks (e.g., TFLite [65] or MXNet [4]) or platforms (e.g., TPUs) is an interesting avenue for future work. Although so far we have only applied IvySyn on DL-framework code, we anticipate our approach to generalize to other, non-DL codebases as well, provided there is a synergy between high- and low-level, native (interfacing) APIs.

## 6 Prototype Implementation

IvySyn consists of ≈1.9 KLOC of C/C++ and ≈1.1 KLOC of Python, along with various shell scripts (≈100LOC).
**Instrumentation.** The instrumentation component of IvySyn (for source code rewriting in order to inject the fuzzer code; Figure 1) is implemented using Clang v11.0.1. It consists of ≈900 LOC as follows: in the case of PyTorch, the instrumentation is done by a Python script, which leverages the Clang Python bindings for visiting and instrumenting the target native functions (≈550 LOC of Python); in TensorFlow, we implemented the instrumentation as a native Clang pass (≈300 LOC of C++), along with ≈40 LOC of shell code for running the pass on the files containing the kernel implementations, and instrumenting everything accordingly. Instrumentation via means of static binary code rewriting via the Egalito framework [73], as well as rewriting via Intel's Pin DBI framework [40], could be supported in the future.
**Native Fuzzer.** The native fuzzer component of IvySyn is implemented in C++ as an *in-tree* part for each framework.

In total, it consists of ≈2.1 KLOC. To incorporate Address-Sanitizer (ASan) in `IvySyn` to support memory inspection and security analysis (§7.5), we modified the build process of PyTorch and TensorFlow. Finally, the Watchdog required to run the developer tests (and to restart them in the case of a crash) was implemented in ≈200 LOC of Python.

**Synthesizer.** The synthesizer component is implemented in ≈750 LOC of Python for parsing the logged crash files and generating the Python PoVs, along with ≈100 LOC of shell code for categorizing the reproduced PoVs.

# 7 Evaluation

In this section, we assess `IvySyn` both in terms of its efficiency in uncovering low-level crashing inputs over time, as well as in terms of its effectiveness in leveraging crashing inputs to synthesize PoVs that trigger the respective errors from high-level (framework) APIs. More specifically, we answer the following research questions:

**RQ1:** Is `IvySyn` efficient in uncovering crashing inputs over time? (§7.2)
**RQ2:** Is `IvySyn` effective in leveraging crashing inputs to synthesize PoVs? (§7.3)
**RQ3:** Which `IvySyn` mutations are the most successful in uncovering memory errors? (§7.4)
**RQ4:** What are the security ramifications of the PoVs synthesized with `IvySyn`? (§7.5)

## 7.1 Experimental Setup

**Statistical Significance.** Due to the non-deterministic nature of fuzzing, results of individual experiments tend to vary across multiple runs. We therefore ran our experiments multiple times and report statistically significant results, according to the guidelines of Klees et al. [37]. Specifically, when applicable, we report medians of runs over time, along with their respective 95% confidence intervals (CIs), as well as the $p$-values of two-sided Mann-Whitney U-tests. We also report Cohen's $d$-effect sizes.

**`Atheris`.** We compared `IvySyn`'s efficiency in uncovering crashing inputs against `Atheris` [23], a state-of-the-art *coverage-guided* Python fuzzer that utilizes both Python and native (C/C++) code coverage information. `Atheris` is a production-grade tool, built by Google, and is used to fuzz test their TensorFlow codebase. However, `Atheris` is not fully automated: it requires helper code (i.e., drivers) to convert an input byte-buffer into the proper Python API arguments [24]. Next, it uses libFuzzer [55] to perform coverage-guided mutations on its input byte-buffer and derive values (i.e., series of raw bytes) that, when turned into Python API arguments, will increase both native and Python code coverage. We used `IvySyn` to generate `Atheris` drivers for TensorFlow and PyTorch kernels, as follows.

First, we ran `IvySyn` and used its argument logging functionality to collect the argument types of kernels APIs. Next, we used `IvySyn`'s synthesizer to derive the respective high-level API arguments and constructed code snippets involving the corresponding Python APIs. Then, we used these snippets to generate drivers for two variants of `Atheris`, called `Atheris+` and `Atheris++`, to compare `IvySyn` against `Atheris` in terms of efficiency. We did not compare against `Atheris` in terms of effectiveness in synthesizing PoVs, because `Atheris` lacks such functionality.

• `Atheris+` *(drivers without type awareness)*: We provided drivers that invoke the Python APIs with the correct number of arguments, but let `Atheris` choose, from its default pool (`tf.bool`, `tf.string`, `tf.int32`, *etc.*), the type that will increase code coverage when assigned to each argument. This variant is similar to traditional byte-level mutation-based fuzzers (e.g., AFL), and corresponds the *default* operation of the tool. However, without `IvySyn`'s automation, putting together such drivers still requires manual effort.

• `Atheris++` *(drivers with type awareness)*: This is not an off-the-shelf tool. Instead, it borrows automation for driver generation and type information, from `IvySyn`, regarding arguments of type `string`, `bool`, numerical, and tensor. We built this variant to compare `IvySyn` against a stronger baseline, which leverages type information, as well as native and Python code-coverage feedback.

**`DocTer`.** We compared `IvySyn`'s effectiveness in synthesizing PoVs against `DocTer` [74]. Similarly to `IvySyn`, `DocTer` also performs mutations on high-level APIs, and generates test cases that may trigger fatal runtime and memory errors on kernel implementations. `DocTer`'s test cases are therefore directly comparable to `IvySyn`'s PoVs. However, we did not compare in terms of efficiency against `DocTer`, because the later is not fully automated. Instead, it requires a domain-expert to add annotations (specifying valid argument-value combinations for target APIs) and it is unclear how to reason about the time required for his manual effort. Yet, we reused existing annotations and the `DocTer` authors' instructions to compare the number of PoVs/test cases generated by both tools, on a subset of the latest TensorFlow and PyTorch APIs, for which the available annotations were directly reusable.

**Target Frameworks.** We present results obtained on TensorFlow v2.6 (commit `919f693`) and PyTorch v1.11 (commit `bc2c6eda`). `IvySyn` automatically instrumented 1440 TensorFlow and PyTorch kernels with CPU or GPU implementations, or both, and eventually fuzzed 1159 of those [33].

• *TensorFlow*: `IvySyn` automatically instrumented 539 of the 669 TensorFlow kernels inheriting from the `OpKernel` class, excluding 80 kernels with shared state (i.e., kernels which either include a `mutex` or interact with a `ResourceMgr` object), and 50 edge cases, such as `Compute` APIs with an empty body or non-standard signatures (§5.1). Out of the 539 instrumented TensorFlow kernels, 412 were reachable by developer tests, and these were eventually fuzzed by `IvySyn`.

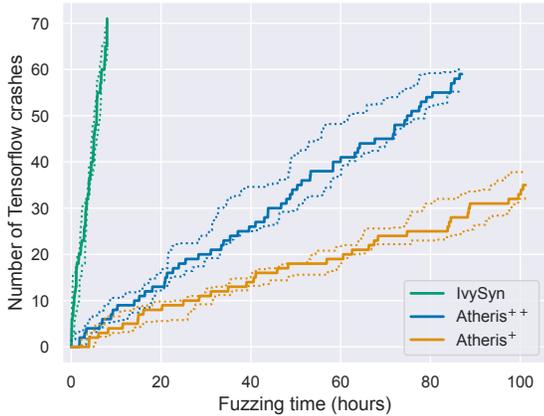

Figure 2: Total number of crashes found by `IvySyn` vs. `Atheris` in 308 TensorFlow kernels over time. Continuous lines correspond to medians over 5 iterations; dotted lines show the 95% CIs ($p$-values $< 10^{-5}$; Cohen's $d$-effect $> 2$).

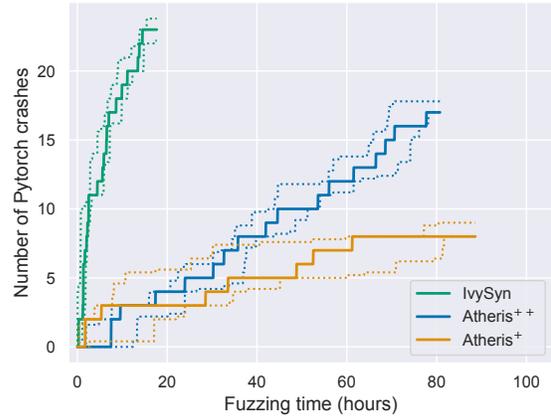

Figure 3: Total number of crashes found by `IvySyn` vs. `Atheris` in 283 PyTorch kernels over time. Continuous lines correspond medians over 5 iterations; dotted lines show the respective 95% CIs ($p$-values $< 10^{-5}$; Cohen's $d$-effect $> 2$).

- *PyTorch*: `IvySyn` automatically instrumented 901 of the 1669 implemented PyTorch kernels (according to `native_functions.yaml`, which contains declarations for the natively-implemented kernels), excluding 768 kernels. A subset of these kernels were excluded because they provide duplicate (or equivalent) functionality to kernels which `IvySyn` already instrumented: 111 so-called `out`-variants of kernels, 73 kernels for in-place operations, and 92 other edge cases, such as helper APIs. `IvySyn` did not instrument 181 kernels with unsupported argument types, 19 kernels for complex, recursive neural network operations, and 292 APIs with no matching declarations. (These kernels could be supported with additional engineering effort [32].) Out of the 901 instrumented PyTorch kernels, 747 were reachable by developer tests, and these were eventually fuzzed by `IvySyn`.

**Timeouts and Hardware Testbed.** We configured all fuzzing sessions to fuzz the respective kernels individually, and in a random order, for up to a 20-min. timeout For each fuzzing session, we used one (user-space) thread and 16GB of RAM, on different (but identically-configured) VMs, with different pseudo-random generator seeds. All fuzzing sessions were completely independent from each other, and fuzzing of individual kernels that did complete in the 20-min. time chunk, either uncovered a crash or exceeded the 16GB of assigned memory; or, in the case of `IvySyn` and `DocTer` only, there were no more mutations left to perform. In all experiments, we used Debian v11 on 4-core Intel Xeon 3.7GHz CPU.

## 7.2 Uncovering Crashing Inputs (RQ1)

To assess the *efficiency* of `IvySyn` in terms of uncovering crashing inputs on native kernel implementations—i.e., *"How quickly can `IvySyn` find crashing inputs?"*—we compare against `Atheris` on TensorFlow and PyTorch. We focused on a subset of 308 TensorFlow and 283 PyTorch kernels (out of the 1159 total fuzzed kernels) for two reasons.

First, `IvySyn` must be able to successfully synthesize a Python snippet for them because `Atheris`, by itself, is not fully automated and relies on the snippets synthesized by `IvySyn` to generate its drivers. Second, the corresponding binding invoked by the synthesized Python code snippet needs to map to exactly one kernel, in order for `Atheris` to spent its 20-min. slot on fuzzing that single kernel, like `IvySyn` does—as opposed to, say, selecting bindings that map to multiple kernels, which may (depending on the input arguments) invoke multiple native kernels, and thus result in less than 20-min. of fuzzing time per kernel.

Figure 2 and Figure 3 show the number of crashes uncovered by `IvySyn`, `Atheris`[+], and `Atheris`[++], over time on TensorFlow and PyTorch kernels, respectively. All three tools uncovered crashing inputs that led to abnormal process termination with `SIGABRT`, `SIGFPE`, or `SIGSEGV` exceptions. Continuous lines show medians of 5 iterations, whereas the dotted lines show 95% CIs. The respective $p$-values were less than $10^{-5}$ and Cohen's $d$-effect sizes were greater than 2 (i.e., they are considered "huge" in the literature).

The difference between the median number of crashes found by `IvySyn` versus the two `Atheris` variants is significant. Not only did `IvySyn` ultimately find *more* crashes in both frameworks (i.e., 71 vs. 59 and 35 in TensorFlow; and 23 vs. 17 and 8 in PyTorch), but it also did so considerably *faster*. For example, in TensorFlow, `IvySyn` found a median of 71 crashes in less than 10 hours, whereas `Atheris`[+] and `Atheris`[++] found a median of 5 and 8 crashes, respectively, in the same time frame; and took more than 9x of total time to complete (i.e., `Atheris`[++] ≈85 hours vs. `IvySyn` ≈9 hours).

| | Fuzzer | TensorFlow | PyTorch |
|---|---|---|---|
| **Total Crashes** | Atheris[+] | 47 | 9 |
| | Atheris[++] | 64 | 18 |
| | IvySyn | **80** | **25** |
| **Union** | All | 87 | 30 |

Table 1: Crashes found by IvySyn and Atheris in 308 TensorFlow and 283 PyTorch kernels. (Aggregate results of 5 iterations for each fuzzer on each framework.)

| Framework | Fuzzed Kernels | Unique Crashes | Synthesized PoVs |
|---|---|---|---|
| **TensorFlow** | 412 | 103 | 86 / 103 (83%) |
| **PyTorch** | 747 | 81 | 49 / 81 (60%) |
| **All** | 1159 | 184 | **135 / 184 (73%)** |

Table 2: PoVs successfully synthesized by IvySyn in TensorFlow and PyTorch, along with the number of fuzzed kernels and the respective number of unique crashes found.

Similarly, in PyTorch, IvySyn found a median of 23 crashes in less than 18 hours, whereas Atheris[+] and Atheris[++] found a median of 3 crashes in the same time frame; and took more than 5x of total time to complete (i.e., Atheris[++] ≈81 hours vs. IvySyn ≈16 hours). Noticeably, in both frameworks, Atheris[++], which leveraged IvySyn-provided type awareness, eventually outperformed Atheris[+] in a statistically significant manner, but fell short compared to IvySyn. Yet, Atheris[++] is not an off-the-shelf tool. Instead, it is a customized, strong baseline that borrows type information from IvySyn. Thus, it was expected to be closer to IvySyn—in terms of total crashes found.

Since Figure 2 and Figure 3 only show the number of crashes found on completely independent fuzzing sessions, Table 1 sheds some light on how the sets of total crashes found by each fuzzer compare across each other. In TensorFlow, IvySyn found 80 crashes in 308 kernels, while Atheris[++] found 64 crashes and Atheris[+] found 47 crashes. Furthermore, the union of crashes of all fuzzers in the 308 TensorFlow kernels was 87. That is, since IvySyn alone found 80 crashes, there were only 7 crashes found by Atheris, but not by IvySyn. In particular, 4 crashes were triggered by Atheris[+] by passing incorrect argument types (IvySyn and Atheris[++] failed to trigger those, because they always provide proper argument types). The remaining 3 crashes were triggered by Atheris[++], but not IvySyn, because the former generated inputs with value combinations that happen not to be in the mutation pools of IvySyn (see §5.3).

In PyTorch, IvySyn found 25 crashes in 287 kernels, while Atheris[++] found 18 crashes and Atheris[+] found 9 crashes. Additionally, the union of all crashes found in the 387 PyTorch kernels was 30. That is, there were only 5 crashes found by Atheris, but not by IvySyn. Similarly to TensorFlow, these 5 crashes were only triggered by Atheris[++] because of crashing inputs not in the mutation pools of IvySyn.

Overall, Figure 2, Figure 3, and Table 1, suggest that type awareness, as introduced by IvySyn, consistently helped both IvySyn and Atheris[++] to (i) find *more* crashing inputs than Atheris[+], and (ii) find these crashing inputs *faster*. Yet, Atheris[++]—a state-of-the-art coverage-guided fuzzer, enhanced with type awareness and automation by us—still fell short compared to IvySyn.

This is because code coverage information rarely helped derive crashing inputs whose values were not in the mutation pools of IvySyn; and even when it did so, it was considerably slower compared to IvySyn.

### 7.3 Synthesizing PoVs (**RQ2**)

To assess the *effectiveness* of IvySyn in terms of synthesizing PoVs, we investigated how well IvySyn can utilize the low-level crashing inputs uncovered during its fuzzing sessions to synthesize high-level code snippets.

Table 2 shows the collective numbers of TensorFlow and PyTorch kernels fuzzed across all our experiments with IvySyn, the number of unique crashes uncovered, and the number of PoVs successfully synthesized from them. In total, IvySyn fuzzed 1159 TensorFlow and PyTorch kernels with either CPU or GPU implementations, or both, and found crashes in 184 of them. Then, armed with those 184 crashes, IvySyn successfully synthesized 135 PoVs (73%), in less than three seconds. IvySyn failed to synthesize PoVs for 17 TensorFlow crashes (17%), and 32 PyTorch crashes (40%), due to current IvySyn limitations. In TensorFlow, out of the 17 crashing kernels for which IvySyn's synthesizer failed to construct PoVs: 2 were deprecated at the Python level; 6 were not directly exposed to Python; 5 kernels required complex list input arguments, which the synthesizer failed to infer; and 4 were not reproducible when installing the pre-built binaries from the PyPI Python package repository. In PyTorch, out of the 32 crashing kernels for which the synthesizer failed to construct PoVs: 31 had no (Python) binding, while for the last case, IvySyn failed to generate a PoV because of complex input(s) required to properly invoke the crashing kernel.

Despite its current limitations, IvySyn helped TensorFlow and PyTorch framework developers identify and fix 61 previously-unknown vulnerabilities, and assign 39 CVEs.

**IvySyn vs. DocTer.** DocTer's test cases also trigger fatal runtime and memory errors, and are comparable to IvySyn's PoVs (§7.1). We modified DocTer to target the specific subset of 125 TensorFlow and 105 PyTorch kernels, which (i) were fuzzed by IvySyn, and (ii) the available DocTer annotations were directly reusable (i.e., the latest APIs maintained the same type signature).

| IvySyn Type-aware Mutations | SIGSEGV∓ TensorFlow | PyTorch | SIGABRT† TensorFlow | PyTorch | SIGFPE TensorFlow | PyTorch | Total PoVs |
|---|---|---|---|---|---|---|---|
| Tensor with random dimension sizes | 21 | 8 | 10 | – | 3 | 4 | 46$^\vee$ |
| Tensors with extreme values | 2 | – | 22 | – | 1 | – | 25$^\pm$ |
| Permutations of original arguments | 1 | 1 | 17 | – | – | – | 19* |
| Zero values | N/A | – | N/A | – | N/A | 12 | 12 |
| Lists with extreme values | N/A | 13 | N/A | – | N/A | – | 13 |
| Tensors with empty shape | 1 | 2 | 5 | – | – | – | 8$^\oplus$ |
| Extreme values in primitive types | N/A | 6 | N/A | – | N/A | – | 6 |
| Empty lists | N/A | 3 | N/A | – | N/A | – | 3 |
| Deep tensors | 1 | – | 2 | – | – | – | 3 |
| **Total PoVs** | 26 | 33 | 56 | 0 | 4 | 16 | 135 |

Table 3: Total number of PoVs per mutation and per crash type. The 39 CVEs of previously-unknown security vulnerabilities uncovered by IvySyn are distributed as follows: 21$^\vee$, 8$^\pm$, 7*, and 3$^\oplus$ (per mutation type); and 17$^\mp$ and 22$^\dagger$ (per crash type).

We run DocTer on each framework, according to its developers' instructions and default settings, and compare with IvySyn over 10 runs. In TensorFlow, the median of successfully synthesized PoVs over 10 iterations was 15 PoVs (union: 19) for IvySyn vs. 12 PoVs (union: 16) for DocTer. Furthermore, the average total running time was 186 mins (std. dev. 5 mins) for IvySyn vs. 197 mins (std. dev. 6 mins) for DocTer. In PyTorch, the median of successfully synthesized PoVs over 10 iterations was 11 PoVs (union: 14) for IvySyn vs. 7 PoVs (union: 9) for DocTer. Additionally, the average total running time was 560 mins (std. dev. 20 mins) for IvySyn vs. 738 mins (std. dev. 14 mins) for DocTer.

In both frameworks, IvySyn consistently found more PoVs than DocTer with $p$-values less than $10^{-2}$ and Cohen's $d$-effect sizes greater than 2. Across all runs combined, IvySyn found 10 TensorFlow and 9 PyTorch PoVs, which DocTer did not manage to find. Conversely, DocTer found 7 TensorFlow and 4 PyTorch PoVs, which IvySyn failed to discover. IvySyn failed to derive these PoVs for two main reasons. First, as described in Section 5.1, IvySyn fuzzes inputs provided to the Compute function of the OpKernel class. However, some of the high-level TensorFlow API parameters correspond to static attributes in the OpKernel class, which are not passed as inputs to Compute; these attributes are thus fuzzed by DocTer, but not IvySyn. Second, the two tools follow different strategies for filling tensors with values: DocTer fills in random values uniformly, whereas IvySyn fills tensors with one value. We leave as future work the possibility of further improving IvySyn's implementation, inspired by DocTer.

Overall, IvySyn's total running time was consistently less than DocTer's in both frameworks, and IvySyn consistently found more PoVs in both frameworks. Ultimately, although DocTer leveraged domain-expert annotations—and it is unclear how to reason about the total time needed to compile them—during the course of comparable, fully-automated runs, IvySyn found more PoVs than DocTer per unit of time.

### 7.4 Type-aware Mutations (RQ3)

In Section 5.3, we empirically studied CVEs assigned to TensorFlow vulnerabilities and PyTorch bug reports that involve memory safety issues, and compiled a set of type-aware mutations relevant to DL operations. Furthermore, in RQ1 we already showed that, by iterating over its type-aware mutation pools, IvySyn significantly outperforms approaches that leverage code coverage information to derive test inputs.

We now investigate exactly how many PoVs were due to each IvySyn type-aware mutation category, and what were the specific crash types triggered. Table 3 shows, in descending order, the total number of PoVs per mutation type, as well as the respective crash types triggered on each framework. Cases where the specific mutation type was not applicable on the target framework or no crashes/PoVs were uncovered, are marked as $N/A$ and empty cells, respectively.

We make the following observations. First, mutations that involved tensors with random dimensions uncovered the greatest number of PoVs: 46 in total (34 in TensorFlow, 12 in PyTorch). The specific crash types triggered were SIGSEGV (21 in TensorFlow, 8 in PyTorch), SIGABRT (10 in TensorFlow), and SIGFPE (3 in TensorFlow, 4 in PyTorch). Second, mutations involving tensors consisting of extreme values led to 25 PoVs, all in TensorFlow. The specific crash types were 2 SIGABRT, 22 SIGFPE, and 1 SIGSEGV. Third, permutations of original argument values led to 18 TensorFlow and 1 PyTorch PoVs, and the respective crash types were SIGSEGV (1 in TensorFlow, 1 in PyTorch) and SIGABRT (17 in TensorFlow). Fourth, mutations involving zero values led to 12 PyTorch PoVs (all SIGFPE). Fifth, mutations involving lists consisting of extreme values led to 13 PyTorch PoVs (all SIGSEGV).

Finally, regarding the aggregate number of crash types, IvySyn's type-aware mutations uncovered 26 and 33 SIGSEGV, and 4 and 16 SIGFPE, in TensorFlow and PyTorch, respectively; but 56 versus 0 SIGABRT.

This is because in the codebase of TensorFlow, `SIGABRT` signals are explicitly raised in case of sanity checks failing unexpectedly. In contrast, PyTorch does not raise such signals.

Overall, by utilizing various categories of type-aware mutations, `IvySyn` uncovered 135 PoVs across two different frameworks. These PoVs, and the respective bug reports, were perceived critically by the corresponding developers, who confirmed and fixed (with high-priority) 61 previously-unknown security vulnerabilities, and assigned 39 CVEs.

## 7.5 Security Analysis (RQ4)

`IvySyn` successfully synthesized PoVs for 135 of the total 184 TensorFlow and PyTorch crashes found. These PoVs triggered abort signals (i.e., `SIGABRT`), floating-point exceptions (i.e., `SIGFPE`), and segfaults (i.e., `SIGSEGV`), and halted the respective execution. To assess how, and to what extent, these PoVs could be abused to corrupt memory locations, leak program contents, or cause the respective runtime environment to crash, we conducted a security analysis using ASan.

In TensorFlow, `IvySyn` synthesized PoVs for 4 floating-point exceptions, 56 abort signals, and 26 segfaults. In PyTorch, `IvySyn` synthesized PoVs for 16 floating-point exceptions and 33 segfaults. Although floating-point exceptions and aborts do not cause memory corruption, they are still a security-relevant threat, as they can crash the DL framework runtime and cause a DoS. Overall, segfaults are more severe, and we focused our analysis on them. Specifically, we used ASan to categorize the 26 TensorFlow and 29 out of the 33 PyTorch segfaults according to their root cause. The remaining 4 PyTorch segfaults were in GPU-specific kernel implementations, thus they could not be analyzed by ASan.

In TensorFlow, 18 out of the 26 segfaults were caused by a `NULL`-pointer dereference (3 memory-write and 15 -read operations). These 18 crashes, similarly to the floating-point exceptions and aborts, cannot be directly leveraged by an attacker to gain a read/write primitive [58]. Yet, they could still be used to crash the DL framework and facilitate a DoS attack. The remaining 8 segfaults were reported by ASan as memory read (6) and write (1) operations at controlled addresses, and 1 memory write caused by a heap overflow. By controlling the addresses of these operations, an attacker could leverage them to gain arbitrary read/write primitives [48].

In PyTorch, 10 out of the 29 segfaults were caused by a `NULL`-pointer dereference (3 memory-write and 7 -read operations). The remaining 19 segfaults were security-critical: 3 were caused by an out-of-bounds memory write operation on the heap—an attacker could carefully massage the heap and leverage such bugs to gain a write primitive on objects next to the vulnerable one; 12 segfaults were reported by ASan as memory read operations at controlled addresses, and 4 segfaults were reported as memory read operations at heap addresses. Again, by controlling the addresses of such operations, an attacker could gain arbitrary read/write primitives.

## 8 Related Work

Our work aims at testing DL frameworks through their developer-accessible APIs using fuzzing, and relates with various techniques, including mutation-based [6, 55] and grammar-based [7, 68] fuzzing, model-based testing [5, 11, 13, 19, 27], and learning-based testing [3, 7]. These techniques have been extensively used in the past for program testing in binary and non-binary input domains, as well as for API testing across domains, ranging from OS kernel drivers to cloud services APIs and DL framework APIs.

**Binary-input Testing.** Random input fuzzing was first introduced in the 90's in an attempt to automatically find reliability issues in Unix command-line utilities [41]. Most recent mutation-based fuzzing tools, like AFL and libFuzzer [55], are guided by code-coverage feedback and have found hundreds of CVEs. Typically, coverage-guided fuzzing tools are further improved with search strategies that specifically target rare code paths [9], avoid path collisions [20], target specific program locations efficiently [8], and leverage neural networks in order to flip difficult-to-pass branches [10, 56].

`IvySyn` is similar to the above works in that it also applies mutation-based fuzzing to uncover memory errors in C/C++ code. However, unlike `IvySyn` that attempts to find crashes in APIs of DL frameworks, all the above works target domains with binary-input formats, such as audio and image processing applications, ELF parsers, and command-line utilities.

**Non-binary-input Testing.** In domains with complex, non-binary input structure—e.g., XML parsers, compilers, and interpreters—traditional mutation-based fuzzing has been combined with grammar-aware mutations [45, 57, 69] in order to help generate test cases that reach execution stages beyond syntactic or semantic checks. Furthermore, grammar-based fuzzing—using probabilistic context-sensitive grammars [68], context-free grammars [7], or automata [3]—has also been used to synthesize test cases for non-binary input domains.

Although `IvySyn` contains a synthesis component to ultimately generate PoVs, its synthesizer does not require grammars. Instead, it follows a bottom-up approach: starting from a given set of known crashing inputs in a low-level language, it synthesizes a valid program in a managed language, which, when executed, will propagate the specific offending inputs to the native APIs. Our synthesis approach is straightforward (as the cross-language mapping between low- and high-level APIs is inherent to the design of DL frameworks), and requires only few seconds to synthesize tens of PoVs in a complex language (e.g., Python).

**Testing System Interfaces.** `IvySyn` is related to past works that focus on finding errors in interface implementations of systems in various domains, ranging from testing cloud service APIs [5, 19] to OS kernel interfaces [11, 13, 27] and native library interfaces [6]. Similarly to the above, `IvySyn` also takes into consideration the additional notion of calling context: that is, invoking an API requires not only the proper argument types, but also the proper caller(s) chain and global state (i.e., variables initialization). However, instead of leveraging, say, model-based testing [5, 11, 13, 19, 27] (to infer dependencies and generate API sequences), `IvySyn` relies on developer unit tests to force-execute the target APIs, which are then invoked and fuzzed with a proper calling context.

Furthermore, unlike all past works that automatically synthesize fuzzing drivers [6], `IvySyn` is performing mutation-based fuzzing on native code and uses crashing inputs to synthesize PoVs on a high-level, managed language. Thus, `IvySyn` operates across layers of abstractions implemented in different low- and high-level languages, with different security assumptions (instead on a homogeneous C/C++ stack).

**Testing Deep Learning Frameworks.** `IvySyn` most closely relates to past works that aim at finding bugs in the underlying implementations of DL frameworks. Mutation-based approaches that make use of pre-trained DL models as seeds [18, 26, 44, 71, 72, 75] mostly uncover errors with regards to incorrect or inconsistent functionality, numerical errors, and performance degradation; while others that perform mutations on the input parameters of high-level DL framework APIs uncover—similarly to `IvySyn`—fatal runtime and memory safety errors in DL kernel implementations [23, 74].

However, unlike `IvySyn`, these tools are not fully-automated. Specifically, `DocTer` [74] requires a domain-expert to manually add annotations for specifying valid argument-value combinations. Likewise, in `Atheris` [23], a developer has to manually implement driver code and add type-awareness with respect to high-level APIs.

Conceptually, `IvySyn`'s main difference compared to prior work in testing DL frameworks is the design and implementation of a two-fold, bottom-up approach. To the best of our knowledge, `IvySyn` is the first fuzz testing framework to realize such an approach and has three key benefits: (1) it is fully-automated and requires no manual effort, like manually constructing fuzzing drivers or adding domain-expert annotations; (2) it hooks directly on the strongly-typed native APIs and can perform type-aware mutations, targeted to test DL framework APIs; and (3) it leverages inherent DL framework mappings and low-level crashing inputs to seamlessly synthesize code snippets involving the respective high-level APIs. As shown earlier (§7), these benefits allow `IvySyn` to automatically find more crashing inputs than low-level fuzzers, and more PoVs than comparable high-level API fuzzers.

**Testing Deep Learning Models.** Lastly, a target domain related, but orthogonal, to ours is testing DL models with respect to their robustness against adversarial examples [22]. Adversarial examples that confuse DL models is a topic of intense research with attacks [25, 47, 75] and defenses [12, 39] developed in an iterative manner. Our target domain relates with that of the above works, but our goal is orthogonal since our objective is to uncover memory errors in the implementations of DL frameworks; not bugs in pre-trained DL models in order to improve robustness against adversarial examples.

## 9 Conclusion

DL frameworks involve code that spans different low- and high-level languages. In this context, it is no surprise that, due to missing sanity checks and mismatched security assumptions, untrusted inputs may transfer through the stack of APIs and reach memory-unsafe code. The question then becomes: *"How can we automatically detect such code defects and report them to framework developers in an actionable manner?"*

With `IvySyn`, we presented a path towards answering this question. Specifically, `IvySyn` first identifies DL kernel code implementations and adds fuzzing hooks in order to perform mutation-based fuzzing with type-aware mutations. Next, given a set of crashing kernels, it synthesizes high-level code snippets that propagate offending inputs, which crash native kernels, thought high-level APIs. Such code snippets serve as a proof of vulnerability, and can help developers replicate, analyze, and eventually fix the respective bug(s).

Although still an early prototype, `IvySyn` has already helped TensorFlow and PyTorch framework developers *identify and fix 61 previously-unknown security vulnerabilities, and assign 39 unique CVEs*. Arguably, `IvySyn` has improved the security posture of the two most popular DL frameworks, explicitly and implicitly affecting millions, if not billions, of users worldwide. We posit that in such a rapidly evolving ecosystem, the need for fully-automated vulnerability discovery frameworks, like `IvySyn`, is undeniable.

## Availability

The prototype implementation of `IvySyn` is available at: https://gitlab.com/brown-ssl/ivysyn

## Responsible Disclosure

The authors followed the community guidelines [49, 64] and responsibly disclosed the discovered vulnerabilities, and bugs, to the framework developers of PyTorch and TensorFlow.

## Acknowledgments

We thank our anonymous shepherd and reviewers for their valuable feedback. This work was supported in part by the CIFellows 2020 program, through award CIF2020-BU-04, and the National Science Foundation (NSF), through awards CCF-2107405 and CCF-1845893. Any opinions, findings, and conclusions or recommendations expressed herein are those of the authors and do not necessarily reflect the views of the US government, NSF, or CRA.

## A  Instrumentation Example

`IvySyn` automatically instruments DL kernel implementations and injects wrappers that kick-start a force-executed fuzzing session. Listing 3 shows an example of such a wrapper for the PyTorch kernel `cosine_similarity`.

```cpp
Tensor cosine_similarity(const Tensor& x1,
    const Tensor& x2, int64_t dim, double eps){

    if (fuzzing::already_fuzzing ||
        fuzzing::was_fuzzed("cosine_similarity"))
        // Return the original function result
        return do_cosine_similarity(x1, x2,
            dim, eps);
    fuzzing::already_fuzzing = true;

    Tensor retval, x1_fuzz, x2_fuzz;
    int64_t dim_fuzz; double eps_fuzz;
    std::vector<std::string> types =
        {"Tensor&", "Tensor&",
         "int64_t", "double"};
    std::vector<void *> args{};
    args.push_back((void *) &x1);
    ... // Repeat for x2, dim, and eps
    // Initialize fuzzer with the original
    // args and types
    fuzzing::Fuzzer fuzzer =
        fuzzing::Fuzzer("cosine_similarity",
            types, args);

    while (fuzzer.has_more_mutations(true)){
        // Get the next combination of inputs
        x1_fuzz = fuzzer.get_next_mut_tensor();
        ... // Repeat for x2, dim, and eps
        try {
            // Invoke the original function
            fuzzer.mut_start_time();
            do_cosine_similarity(x1_fuzz,
                x2_fuzz, dim_fuzz, eps_fuzz);
            fuzzer.mut_end_time(false);
        } catch (...) { ... }
    }
    fuzzing::already_fuzzing = false;

    // Return the original function result
    return do_cosine_similarity(x1, x2,
        dim, eps);
}
```

Listing 3: Example instrumentation performed by `IvySyn` on the kernel `cosine_similarity` of the PyTorch API `torch.nn.functional.cosine_similarity`.

## B  Root-cause Analysis

To provide a better insight regarding how the offending inputs discovered by the PoVs of `IvySyn` can lead to memory or fatal runtime errors, we discuss the root causes of three crashes in kernels across PyTorch and TensorFlow. All three cases were previously-unknown vulnerabilities, which were first discovered by `IvySyn`, while the two TensorFlow vulnerabilities presented were also assigned CVEs [15, 16].

### B.1  `_remove_batch_dim` (PyTorch)

`IvySyn` uncovered a memory corruption vulnerability in PyTorch's `_remove_batch_dim` [53] kernel; Listing 4 shows

```
1   Tensor _remove_batch_dim(const Tensor& self,
2       int64_t level, int64_t batch_size,
3       int64_t out_dim) {
4     if (!has_level(self, level)) {
5       auto self_sizes = self.sizes();
6       VmapDimVector expanded_sizes(
7           self_sizes.begin(), self_sizes.end());
8       expanded_sizes.insert(
9           expanded_sizes.begin() + out_dim,
10          batch_size);
11      return self.expand(expanded_sizes);
12    // ...
13  }
```

Listing 4: Relevant code from the vulnerable `_remove_batch_dim` kernel in PyTorch.

```
1   class EditDistanceOp : public OpKernel {
2     void Compute(OpKernelContext* ctx) override {
3       // ...
4       if (g_truth == g_hypothesis) {
5       auto loc = std::inner_product(g_truth.begin(),
6           g_truth.end(), output_strides.begin(),
7           int64_t{0});
8       OP_REQUIRES(
9           ctx,
10          0 <= loc &&
11          loc < output_elements,
12          errors::Internal("..."))
13      output_t(loc) =
14          gtl::LevenshteinDistance<T>(truth_seq,
15            hypothesis_seq, cmp);
16      // ...
17  }
```

Listing 5: Relevant code from the vulnerable `Compute()` function of the `EditDistanceOp` kernel in `TensorFlow`. The highlighted line indicates the bug-fixing patch, which was omitted in the original version of the kernel.

the relevant code. In particular, when `has_level` returns `false`, execution will reach ln. 6, where the *attacker-provided* value `batch_size` will be written at an offset from the `expanded_sizes` vector, which is also determined by the *attacker-provided* value `out_dim` (ln. 8–10). Since there are no sanity checks on either argument, and both are 64-bit values, the attacker can essentially gain an 8-byte, *controlled* ("buffer underflow") write and further tamper-with the control- or data-flow of the respective process [70].

### B.2 `EditDistanceOp` (TF)

`IvySyn` uncovered a memory corruption in TensorFlow's `EditDistanceOp` kernel; Listing 5 shows the relevant code. The `loc` variable is computed as the inner product of the *attacker-controlled* `g_truth` and `output_strides` variables (ln. 5-7). This variable is then used as an index for a write operation relative to the `output_t` array.

```
1   void Compute(OpKernelContext* ctx) override {
2     const Tensor& handle = ctx->input(0);
3     OP_REQUIRES(ctx, TensorShapeUtils::IsScalar(
4         handle.shape()),
5         errors::InvalidArgument( "`handle`
6           must be scalar"));
7     const string& name =
8      handle.scalar<tstring>()();
9     // ...
10  }
```

Listing 6: Relevant code from the vulnerable `Compute()` function of the `DeleteSessionTensorOp` kernel in `TensorFlow`. The highlighted lines indicate the bug-fixing patch, which was omitted in the original version of the kernel.

The upper bound of the `loc` value is properly checked (ln. 11), however no check on whether the value is *positive* exists. As a result, an attacker can perform a *controlled* write, in the vicinity of `output_t`, by providing inputs such that the `loc` variable results in a negative value. After reporting this vulnerability to the TensorFlow developers, a check on the value of the `loc` variable was added (ln. 10).

This vulnerability was deemed of *critical severity* and was assigned a *high* CVE score [16], as it allows an attacker to hijack the control- or data-flow of the respective process.

### B.3 `DeleteSessionTensorOp` (TF)

`IvySyn` uncovered a fatal runtime error in TensorFlow's `DeleteSessionTensorOp` kernel. In this case, the attacker can trigger the error by providing a non-1-dimensional tensor as the input for the `handle` argument, resulting in abnormal process termination. As shown in Listing 6 (ln. 6), the `scalar()` method will be invoked on the input tensor. Internally, `scalar()` checks whether the provided tensor is 1-dimensional, and exits the program by calling `abort()` if the check fails. The kernel developers wrongly assumed that the user will always provide a 1-dimensional input and omitted to check the input shape *before* calling `scalar()`. After reporting this vulnerability, it was patched by inserting a check and throwing a Python `InvalidArgument` error (ln. 3-5).

Similar wrong assumptions about the shapes and contents of the input arguments exist across the entire TensorFlow codebase, resulting to a high number of missing validation checks, thereby allowing an attacker cause a DoS. Various other vulnerabilities discovered by `IvySyn` were handled in a similar manner, with the TensorFlow developers adding the proper checks instead of making assumptions about the provided input [60, 62, 63]. Even though an attacker may not gain a memory corruption/disclosure primitive, the promptness in the response of the TensorFlow security team, as well as the assignment of CVEs, indicate the importance of identifying and fixing these types of (DoS) vulnerabilities.

# C Artifact Appendix

## C.1 Abstract

This is the artifact appendix for the `IvySyn` fuzzing framework. It contains instructions about how to setup, run, and reproduce the results of `IvySyn`, along with information regarding system and resource requirements.

## C.2 Description & Requirements

### C.2.1 Security, privacy, and ethical concerns

`IvySyn` is a fuzzer for discovering security-critical bugs in Deep Learning (DL) frameworks. The list of APIs for which `IvySyn` has uncovered bugs, during our experiments, is available in the [repository](#) of the project. If, during the reproduction of any reported result, `IvySyn` produces PoVs for APIs not listed in project repository, contact the authors via HotCRP or follow the *Responsible Disclosure* steps in `README.md` to report the (newly discovered) vulnerabilities to the developers of the corresponding framework(s).

### C.2.2 How to access

`IvySyn` is available at: https://gitlab.com/brown-ssl/ivysyn/-/tree/4b3d26dda0ddea11282c2658e28090a738dfd6c7 (stable ref.)

### C.2.3 Hardware dependencies

The provided Docker images are configured to use 4 CPUs and 16GB of RAM, but can also be set to use fewer (or more) resources, as needed.

### C.2.4 Software dependencies

We provide a Docker image that builds and runs `IvySyn`, and hence [Docker](#) is required. Some scripts run outside Docker containers and were tested on Debian v11—but they are relatively simple and should work on any Linux distribution.

### C.2.5 Benchmarks

All the data required for running our benchmarks are either included in the project repository or can be produced by our scripts during setup. Note that the prototype implementation of `IvySyn` fuzzes both CPU- and GPU-specific implementations of DL kernels. However, we are not able to provide access to machines with GPUs. Therefore, the benchmarks in the artifact fuzz only kernels with CPU implementations. This does not have any effect on the claims of the paper, other than a smaller number of instrumented and fuzzed kernels.

## C.3 Setup

### C.3.1 Installation

To setup `IvySyn`, simply invoke `docker/download-prebuilt-image.sh`. This script will download a pre-built Docker image of `IvySyn`. Alternatively, in order to build the `IvySyn` Docker image from scratch, invoke the script `docker/build-docker-image.sh`, under the root directory of the project repository. (Note that building the image from scratch requires ≈3.5 hours on a 16-core, 64GB RAM host; and the total size of the fully built image is ≈35GB.)

- To start the container, simply run:
  `docker/run-docker-image.sh`
  (This script is configured to run the container with access to 4 CPUs and 16GB of RAM.)

- To provide more/less resources to the container, use the optional `--memory` (e.g., `--memory 8g`) and `--cpus` (e.g., `--cpus 2`) arguments to the `run-docker-image.sh` script. If you increase the number of CPUs the container can use, you should also increase the amount of RAM, since more parallel jobs may require more memory.

- To get a shell on the container, run:
  `docker exec -it ivysyn-instance /bin/bash`

If you wish to run the experiments for comparing `IvySyn` with the two other fuzzers, namely `Atheris` and `DocTer`, an additional ≈39GB of storage is required for their corresponding Docker images (i.e., ≈26GB for `Atheris` and ≈13GB for `DocTer`). For more details, refer to C.4.1 and C.4.2.

### C.3.2 Basic Test

For a quick smoke test, we recommend invoking the `do-run.sh` script, *inside the running Docker container of `IvySyn`*, and providing a small number of kernels to be fuzzed with the `--nkernels` argument. For example:

`./do-run.sh --seed 1 --pytorch --nkernels 5`

Once the script done, it should display a summary of the run, containing information about how to inspect the results.

## C.4 Evaluation workflow

### C.4.1 Major Claims

**(C1):** `IvySyn` *automatically* fuzzes DL frameworks and produces *Proofs of Vulnerability (PoVs)*—i.e., code snippets that trigger memory errors in low-level (C/C++) code of the respective framework via a high-level (Python) API.

**(C2):** `IvySyn`: (i) uncovers *more crashes* in the DL frameworks than the state-of-the-art Python fuzzer `Atheris`, and (ii) it does so *faster*. This is proven by the experiment described in Section 7.2 of our paper.

**(C3):** `IvySyn` produces *more PoVs per unit of time* than `DocTer`, yet another DL-framework fuzzer. This is proven by the experiment in Section 7.3 of our paper.

### C.4.2 Experiments

In what follows, we provide instructions on how to setup, run, and interpret the results of our experiments. All compute-time approximations assume you are using 16GB of RAM and 4 CPUs to run the Docker containers. For additional details, see `README.md` at the root of the project repository.

**(E1):** *[Up to 58 compute-hours – Up to 28 compute-hours with suggested configuration + Up to 2GB disk]: Run `IvySyn` on a selected framework and produce PoVs.*

**Preparation:** Run and connect to the provided (or custom-built) Docker image.

**Execution:** To perform a full run of `IvySyn`, invoke the `do-run.sh` script, *inside the running Docker container of IvySyn*, by providing an integer as the RNG seed and either `--tensorflow` or `--pytorch` to choose the framework to be fuzzed, as follows: `./do-run.sh --seed 123 --tensorflow`

However, note that a full run will require ≈45 hours for PyTorch and ≈12 hours for TensorFlow. We advise restricting the amount of kernels that will be fuzzed, by providing the extra `--nkernels` argument. Furthermore, we suggest fuzzing 300 kernels for each framework. This will require ≈17 hours for PyTorch and ≈9 hours for TensorFlow, while still producing PoVs. Running on TensorFlow will require an additional 2 hours for the first invocation of the script, in order to compile the C++ developer-provided TensorFlow tests, which are also used by IvySyn. Execute `IvySyn` as follows: `./do-run.sh --seed <rng seed> --tensorflow --nkernels 300`

**Results:** Once fuzzing is done, `IvySyn` will produce PoVs under `results/<framework>/ivysyn_povs` (where `<framework>` is `pytorch` or `tensorflow`). To manually run and reproduce the PoVs, do the following (in the `IvySyn` Docker container):

1. Activate the environment of the `pip`-installed version of the corresponding framework. In the case of PyTorch, run: `source /home/ivyuser/ivysyn/venv/anaconda3/bin/activate; conda activate pytorch-1.11-orig`. For TensorFlow, run: `source /home/ivyuser/ivysyn/venv/tensorflow-2.6-orig/bin/activate`.

2. Run the PoVs produced under `/home/ivyuser/ivysyn/results/<framework>/ivysyn_povs` using `python3 <pov.py>` (where `<pov.py>` is the filename of the selected PoV).

**(E2):** *[Atheris-comparison] [Up to 400 compute-hours – Up to 26 compute-hours with suggested configuration + Up to 40GB disk space]: Run `IvySyn` and a selected variant of the `Atheris` fuzzer, and compare their efficiency at uncovering crashing inputs.*

**Preparation:** We provide a separate Docker image that sets-up the two variants of `Atheris`, namely `Atheris+` and `Atheris++`, which are described in Section 7.1 of our paper. Similarly to the `IvySyn` image, you can build this image from scratch, by running:

`comparisons/atheris_comp/docker_env/docker/build-docker-image.sh`

or download a pre-built version of the image by running:

`comparisons/atheris_comp/docker_env/docker/download-prebuilt-image.sh`

**Execution:** To perform the experiment that compares `IvySyn` to `Atheris`, invoke the `compare-fuzzers.sh` script *outside the IvySyn container*.

You need to specify an RNG seed, the framework to fuzz (either `--tensorflow` or `--pytorch`), the `Atheris` variant (`--atheris1`, which corresponds to `Atheris+`; or `--atheris2`, which corresponds to `Atheris++`), and whether you want to limit the number of kernels to be fuzzed using the `--nkernels` argument. For example:

`./compare-fuzzers.sh --tensorflow --atheris2 --seed 123 --nkernels 50`

We suggest running at least 50 kernels to get a decent approximation of the overall results.

The script above will:

1. Instrument the subset of 308 TensorFlow and 283 PyTorch kernels we performed our experiment on (depending on the chosen framework) and re-compile the target framework.

2. Execute `IvySyn` on the target kernels, limiting the fuzzed kernels if `--nkernels` was specified.

3. Run the selected `Atheris` variant on the same subset of fuzzed kernels.

4. Output a summary of the results.

**Results:** The `compare-fuzzers.sh` script will display a summary of the results. Specifically, a new directory will be created at `results/<framework>/atheris_comp/` (where `<framework>` is `pytorch` or `tensorflow`), containing the following:

- results.csv: start/end timestamp, as well as whether a crash was found, for each API (CSV file).
- total_time.txt: total time of the run (text file).
- fuzzer_logs_dir.txt: name of the directory in the Docker container with the raw logs produced by the `IvySyn` fuzzer (text file).

The corresponding CSV that contains similar entries for the `Atheris` run can be found at `comparisons/atheris_comp_fuzzed_<framework>.csv` (where `<framework>` is either `pytorch` or `tensorflow`). The raw `Atheris` logs can be found in the `Atheris` Docker container at `/home/ivyuser/ivysyn-atheris/fuzzer_output`. To connect to the `Atheris` Docker container, in order to manually inspect the logs, run `docker exec -it atheris-instance /bin/bash`.

**(E3):** *[DocTer-comparison] [Up to 34 compute-hours – Up to 26 compute-hours with suggested configuration + Up to 15GB disk space]: Run `IvySyn` and `DocTer` and compare their effectiveness at producing PoVs.*

**Preparation:** We provide a separate Docker image that sets-up `DocTer`. Similarly to the `IvySyn` image, you can either build it from scratch, by running:

```
comparisons/docter_comp/docker_env/docker/build-docker-image.sh
```

or download a pre-built version of the image, by running:

```
comparisons/docter_comp/docker_env/docker/download-prebuilt-image.sh
```

**Execution:** To perform the experiment that compares `IvySyn` to `DocTer`, invoke the `compare-fuzzers.sh` script *outside the IvySyn container*. You need to specify an RNG seed, the framework to fuzz (either `--tensorflow` or `--pytorch`), the `--docter` flag, and whether you want to limit the number of kernels to be fuzzed using the `--nkernels` argument. For example:

```
./compare-fuzzers.sh --tensorflow --docter --seed 123 --nkernels 50
```

We suggest running at least 50 kernels to get a decent approximation of the overall results.

The script above will:

1. Instrument the subset of 125 TensorFlow and 105 PyTorch kernels we performed our experiment on (depending on the chosen framework) and re-compile the target framework.
2. Execute `IvySyn` on the target kernels, limiting the fuzzed kernels if `--nkernels` was specified.
3. Run `DocTer` on the same subset of fuzzed kernels.
4. Output a summary of the results.

**Results:** The `compare-fuzzers.sh` script will display a summary of the results. A new directory with the `IvySyn` results will be created at `results/<framework>/docter_comp/` (where `<framework>` is `pytorch` or `tensorflow`), and will contain files similar to the ones mentioned in the Atheris experiment (i.e., `results.csv`, `total_time.txt`, and `fuzzer_logs_dir.txt`).

The corresponding CSV that contains similar entries for the `DocTer` run can be found at `comparisons/docter_comp_fuzzed_<framework>.csv` (where `<framework>` is `pytorch` or `tensorflow`). The raw `DocTer` logs can be found in the `DocTer` Docker container at `/home/workdir/<framework>`. To connect to the `DocTer` Docker container, in order to manually inspect the logs, run `docker exec -it docter-instance /bin/bash`.

## C.5 Version

Based on the LaTeX template for Artifact Evaluation V20220926. Submission, reviewing and badging methodology followed for the evaluation of this artifact can be found at https://secartifacts.github.io/usenixsec2023/.